\documentstyle[epsfig,aps,preprint,pre]{revtex}

\newcommand{\avk}{\left< k \right>}
\newcommand{\equ}[1]{(\protect\ref{#1})}
\newcommand{\Ri}{R_\infty}
\newcommand{\Fi}{\phi_\infty}
\newcommand{\fluck}{\left< k^2 \right>}

\begin{document}

\title{Epidemic outbreaks in complex  heterogeneous networks}

\author{Yamir Moreno$^*$, Romualdo Pastor-Satorras$^\dagger$ 
  and  Alessandro Vespignani$^*$}

\address{$^*$The Abdus Salam International Centre for Theoretical Physics,
  P.O. Box 586, 
  34100 Trieste, Italy\\
  $^\dagger$Departament de F\'{\i}sica i Enginyeria Nuclear,
  Universitat Polit\`{e}cnica de Catalunya\\
  Campus Nord, M\`{o}dul B4,  08034 Barcelona, Spain   \\
  \vspace*{0.5cm}}
\date\today

\maketitle

\tightenlines

\begin{abstract}
  We present a detailed analytical and numerical study for the
  spreading of infections in complex population networks with acquired
  immunity. We show that the large connectivity fluctuations usually
  found in these networks strengthen considerably the incidence of
  epidemic outbreaks.  Scale-free networks, which are characterized by
  diverging connectivity fluctuations, exhibit the lack of an
  epidemic threshold and always show a finite fraction of infected
  individuals.  This particular weakness, observed also in models
  without immunity, defines a new epidemiological framework
  characterized by a highly heterogeneous response of the system to
  the introduction of infected individuals with different
  connectivity.  The understanding of epidemics in complex networks
  might deliver new insights in the spread of information and diseases
  in biological and technological networks that often appear to be
  characterized by complex heterogeneous architectures.

\end{abstract}

\newpage

\section{Introduction}
The epidemiology of heterogeneous networks has largely benefitted from
the need of understanding the spreading of human sexual diseases in
the complex web of sexual partnership \cite{het84,may87,may88}.
Epidemic modeling considered that population groups can be
characterized in classes having different sexual activity or number of
sexual contacts.  This fact leads to models dealing with heterogeneous
populations which are known to enhance the spread of infections as well
as make them harder to eradicate (for a review see \cite{anderson92}).
In this perspective, a limiting case is represented by the newly
identified classes of complex networks (for a review see
\cite{strog01}).  The highly heterogeneous topology of these networks
is mainly reflected in the small average path lengths among any two
nodes (small-world property) \cite{watts99,amaral}, and in a power law
distribution (scale-free property), $P(k) \sim k^{-2 -\gamma}$, for the
probability that any node has $k$ connections to other nodes
\cite{barab99}.  While regular networks present finite connectivity
fluctuations ($\left<k \right>\simeq\left<k^2 \right>$), scale-free (SF)
networks are a limiting case of heterogeneity where connectivity
fluctuations are diverging if $0<\gamma\leq 1$. In other words, the network
nodes possess a statistically significant probability of having a
virtually unbounded number of connections compared to the average
value.  SF networks find real examples in several technological
networks such as the Internet \cite{falou99,calda00} and the
world-wide-web (WWW) \cite{www99}, as well as in natural systems such
as food-webs, and metabolic or protein networks \cite{strog01}.  The
need to understand the dynamics of information transmission, the error
tolerance \cite{barabasi002,newman00,havlin01} and other properties of
complex networks has therefore called for the study of epidemic
modeling in complex networks
\cite{nw99,moore00,abramson01,pv01a,pv01b}.

A surprising result, originated by the inspection of the
susceptible-infected-susceptible (SIS) model, has shown that the
spread of infections is tremendously strengthened on SF networks
\cite{pv01a,pv01b}. Opposite to standard models, epidemic processes in
these networks do not possess any epidemic threshold below which the
infection cannot produce a major epidemic outbreak or an endemic
state. In principle, SF networks are prone to the persistence of
diseases whatever infective rate they may have.  This feature
reverberates also in the choice of immunization strategies
\cite{het78,may84,psvpro} and changes radically many standard
conclusions on epidemic spreading.  This study appears particularly
relevant in the case of technological networks, for instance for the
spreading of digital viruses in the Internet \cite{pv01a}, and it has
soon been generalized by showing that also the
susceptible-infected-removed (SIR) model show the same absence of
epidemic threshold \cite{lloyd01}. These results highlight the study
of epidemic models in complex networks as potentially relevant also in
human and animal epidemiology \cite{lloyd01}, as confirmed recently by
the experimental observation that the web architecture of sexual 
contacts is best described by a scale-free topology 
in which individuals have widely different connectivities \cite{amaral01}.

In this paper we provide a detailed analytical and numerical study of
the SIR model on two prototype complex networks: the Watts-Strogatz
(WS) model and the Barab\'{a}si -Albert (BA) model.  The first model is a
small-world network with bounded connectivity fluctuations, while the
second one is the prototype example of SF network.  The analytical
approach allows us to recover the total size of the epidemics in an
infinite population, in agreement with earlier estimates
\cite{lloyd01}. We are able to find the analytic expression for the
critical threshold as a function of the moments of the connectivity
distribution and we confirm the absence of any finite threshold for
connectivity distributions $P(k)\sim k^{-2 -\gamma}$ with $0<\gamma\leq 3$.  We obtain
the general analytic expression for the total density of infected
individuals and the epidemic threshold at arbitrary $\gamma$ values. Finite
size network effects can be easily evaluated from the analytic
expressions.  Time evolution and other effects of heterogeneity such
as the relative infection incidence in different connectivity classes
can be predicted.  In order to confirm the analytical findings we
perform large scale numerical simulations on the WS and BA networks.
Numerical results are in perfect agreement with the analytical
predictions and confirm that the interplay of complex networks
topology and epidemic modeling leads to a new and interesting theoretical
framework, whose predictions and implications need to be exhaustively
explored.

During the completion of this paper we became aware of a preprint by
Lloyd and May \cite{lloydsir} which reports a comprehensive study of
the SIR model in scale free networks.  This work extends the
preliminary account provided in Ref.~\cite{lloyd01}

\section{The SIR model} 
\label{sec:biosir}

Our theoretical understanding of epidemic spreading is based on
compartmental models, in which the individuals in the population are
divided in a discrete set of states \cite{anderson92,murray}.  In this
framework, diseases which result in the immunization or death of
infected individuals can be characterized by the classical
susceptible-infected-removed (SIR) model \cite{anderson92,murray}.  In
this model individuals can only exist in three different states:
susceptible (healthy), infected, or removed (immunized or dead).  In a
homogeneous system, the SIR model can be described in terms of the
densities of susceptible, infected, and removed individuals, $S(t)$,
$\rho(t)$, and $R(t)$, respectively, as a function of time. These three
magnitudes are linked through the normalization condition
\begin{equation}
  S(t) + \rho(t) + R(t) =1,
  \label{eq:norm}
\end{equation}
and they obey the
following system of differential equations:
\begin{eqnarray}
  \frac{d S}{d t} &=& - \lambda \overline{k} \rho S, \nonumber \\
  \frac{d \rho}{d t} &=& -\mu \rho + \lambda \overline{k} 
   \rho S,  \label{eq:primiser}\\
  \frac{d R}{d t} &=& \mu \rho. \nonumber 
\end{eqnarray}
These equations can be interpreted as follows: infected individuals
decay into the removed class at a rate $\mu$, while susceptibles
individual become infected at a rate proportional to both the
densities of infected and susceptible individuals. 
Here, $\lambda$ is the microscopic spreading (infection) rate, 
and $\overline{k}$ is the number of contacts per unit time that is 
supposed to be constant for the whole population.
In writing this last term of the equations we 
are assuming the {\em homogeneous
mixing} hypothesis \cite{anderson92}, which asserts that the force
of the infection (the per capita rate of acquisition of the disease by
the susceptible individuals) is proportional to the density of
infectious individuals.  The homogeneous mixing hypothesis is indeed
equivalent to a mean-field treatment of the model, in which one
assumes that the rate of contacts between infectious and susceptibles
is constant, and independent of any possible source of heterogeneity
present in the system.  Another implicit assumption of this model is
that the time scale of the disease is much smaller than the lifespan
of individuals; therefore we do not include in the equations terms
accounting for the birth or natural death of individuals.

The most significant prediction of this model is the presence of a
nonzero epidemic threshold $\lambda_c$ \cite{murray}. 
If the value of $\lambda$ is
above $\lambda_c$, $\lambda>\lambda_c$, the disease 
spreads and infects a finite
fraction of the population. On the other hand, when $\lambda$ is 
below the
threshold, $\lambda<\lambda_c$, the total number of 
infected individuals (the epidemic incidence), $\Ri =
\lim_{t\to\infty} R(t)$, is infinitesimally small in the limit of very large
populations (the so-called thermodynamic limit \cite{marro99}). In
order to see this point, let us consider the set of equations
(\ref{eq:primiser}), in which, without lack of generality, we set
$\mu=1$. Integrating the equation for $S(t)$ with the 
initial conditions
$R(0)=0$ and $S(0)\simeq1$ (i.e., assuming $\rho(0) \simeq 0$, 
a very small initial concentration of infected individuals), 
we obtain
\begin{equation}
S(t) = e^{-\lambda \overline{k} R(t)}.  
\end{equation}
Combining this result with the normalization condition
(\ref{eq:norm}), we observe that the total number of infected
individuals $\Ri$ fulfills the following self-consistent equation:
\begin{equation}
  \Ri = 1 - e^{-\lambda \overline{k}\Ri}.  
\end{equation}
While $\Ri=0$ is always a solution of this equation, in order to have
a nonzero solution the following condition must be fulfilled:
\begin{equation}
    \frac{d}{d \Ri } \left. \left( 1  -   
e^{ -\lambda \overline{k} \Ri} \right) \right|_{\Ri=0} > 1. 
\end{equation}
This condition is equivalent to the constraint $\lambda>\lambda_c$, 
where the epidemic threshold $\lambda_c$ takes 
the value $\lambda_c=\overline{k}~^{-1}$ in 
this particular case. By using a  Taylor expansion at 
$\lambda\simeq\lambda_c$ 
it is then possible to obtain the epidemic incidence behavior 
$\Ri\sim(\lambda -\lambda_c)$ (valid above the epidemic threshold). 
It is worth remarking that the  threshold mechanism is related to 
the basic reproductive rate ${\cal R}_0 \sim \lambda \overline{k} $ 
(not to be confused with removed individuals)
usually considered by epidemiologist.  Only if ${\cal R}_0$ is larger than 
unity the infection can
sustain itself, obviously defining a threshold in the spreading rate
$\lambda$ \cite{epidemics}.
As well, in the language of the physics of nonequilibrium phase 
transition \cite{marro99}, 
the epidemic threshold can be considered as completely 
equivalent to a critical point.  
In analogy with critical phenomena, we can consider $\Ri$ as the
order parameter of a phase transition and $\lambda$ as the tuning
parameter. In particular, it is easy to recognize that the SIR
model is a generalization of the dynamical percolation model, that has
been extensively studied in the context of absorbing-state phase
transitions\cite{marro99}.

\section{The SIR model in complex networks}
\label{sec:complexnetworks}

In order to address the effects of contact heterogeneity in epidemic
spreading, let us consider the SIR model defined on a network with
general connectivity distribution $P(k)$ and a finite average
connectivity $\avk = \sum_k k P(k)$. Each node of the network represents
an individual in its corresponding state (susceptible, infected, or
removed), and each link is a connection along which the infection can
spread. The disease transmission on the network is described in an
effective way: At each time step, each susceptible node is infected
with probability $\lambda$, if it is connected to one or more infected
nodes. At the same time, each infected individual becomes removed with
probability $\mu$, that, without lack of generality, we set equal to
unity.

In order to take into account the heterogeneity induced by the
presence of nodes with different connectivity, we consider the time
evolution of the magnitudes $\rho_k(t)$, $S_k(t)$, and $R_k(t)$, which
are the density of infected, susceptible, and removed nodes of
connectivity $k$ at time $t$, respectively. These variables are
connected by means of the normalization condition
\begin{equation}
  \rho_k(t)+S_k(t)+R_k(t) = 1.
\end{equation}
Global quantities such as the epidemic incidence are therefore expressed
by an average over the various connectivity classes; i.e.  
$R(t)=\sum_k P(k)R_k(t)$.
At the mean-field level, these densities
satisfy the following set of coupled differential equations:
\begin{eqnarray}
  \frac{d \rho_k(t)}{d t} & = & -\rho_k(t) +\lambda k S_k(t) \Theta(t), \label{eq:1}\\
  \frac{d S_k(t)}{d t} & = & - \lambda k S_k(t) \Theta(t), \label{eq:2}\\
  \frac{d R_k(t)}{d t} & = & \rho_k(t). \label{eq:3}
\end{eqnarray}
The factor $\Theta(t)$ represents the probability that any given link
points to an infected site. This quantity can be computed in a
self-consistent way \cite{pv01a}: The probability that a link points
to a node with $s$ links is proportional to $s P(s)$. Thus, the
probability that a randomly chosen link points to an infected node is
given by
\begin{equation}
  \Theta(t) = \frac{\sum_k k P(k) \rho_k(t)}{\sum_s s P(s)}= 
\frac{\sum_k k P(k)
    \rho_k(t)}{\avk}.
  \label{eq:definition}
\end{equation}
In this approximation we are neglecting the connectivity correlations
in the network, i.e., the probability that a link points to an infected
node is considered independent of the connectivity of the node from
which the link is emanating. A more refined approximation would
consider the network correlations as given by the conditional
probability $P(k/k')$ that a node with given connectivity $k'$ is
connected to a node with connectivity $k$ \cite{alexei}. Nevertheless,
as we will see in the next sections, this rather crude approximation
is quite able to give account of many of the properties shown by
computer simulations of the model.

The equations \equ{eq:1}--\equ{eq:3}, combined with the initial
conditions $R_k(0)=0$, $\rho_k(0) = \rho_k^0$, and $S_k(0) = 1- \rho_k^0$
completely define the SIR model on any complex network with
connectivity distribution $P(k)$. 
We will consider in particular the case of a homogeneous
initial distribution of infected nodes, $\rho_k^0 = \rho^0$. In this case,
in the limit $\rho^0 \to 0$, we can substitute 
$\rho_k(0) \simeq 0$ and $S_k(0) \simeq
1$. Under this approximation, Eq.~\equ{eq:2} can be directly
integrated, yielding
\begin{equation}
  S_k(t) =  e^{ -\lambda k \phi(t)}
  \label{eq:4}
\end{equation}
where we have defined the auxiliary function 
\begin{equation}
  \phi(t) = \int_0^t \Theta(t') d t' = \frac{1}{\avk} \sum_k k P(k) R_k(t),
  \label{eq:5}
\end{equation}
and  in the last equality we have made use of the definition
\equ{eq:definition}. It is worth remarking that the above equations
are similar to  those obtained in the case of HIV dynamics in heterogeneous 
populations \cite{may88}

In order to get a closed relation for the total density of 
infected individuals, it results more convenient to focus on the time
evolution of the averaged magnitude $\phi$. To this purpose, let us
compute its time derivative:
\begin{eqnarray}
  \frac{ d \phi(t)}{d t} &=& \frac{1}{\avk} \sum_k k P(k) \rho_k(t) 
  = \frac{1}{\avk} \sum_k k P(k) (1 - R_k(t) - S_k(t))\\
  &=& 1 - \phi(t) -  \frac{1}{\avk} \sum_k k P(k) S_k(t).
\end{eqnarray}
Introducing the obtained time dependence of $S_k(t)$ we are led to the
differential equation for $\phi(t)$
\begin{equation}
  \frac{ d \phi(t)}{d t} = 1 - \phi(t) -  \frac{1}{\avk} \sum_k k P(k) e^{ -\lambda k
    \phi(t)}. \label{eq:generalphi}
\end{equation}
Once solved  Eq.~\equ{eq:generalphi}, we can 
obtain the total epidemic 
incidence  $\Ri$ as a function of 
$\phi_\infty = \lim_{t \to \infty} \phi(t)$. Since
$R_k(\infty) = 1 - S_k(\infty)$, we have
\begin{equation}
  \Ri = \sum_k P(k) \left(1 - e^{ -\lambda k \phi_\infty }\right).
  \label{eq:rigeneral}
\end{equation}
Equations \equ{eq:generalphi} and \equ{eq:rigeneral} constitute thus
an alternative representation of the model, with respect to 
Eqs.~\equ{eq:1}--\equ{eq:3}.

For a general $P(k)$ distribution,  Eq.~\equ{eq:generalphi} cannot
be solved in a closed form. However, we can still get useful
information on the infinite time limit; i.e. at the end of the 
epidemics. Since we have that $\rho_k(\infty)=0$ and consequently  
$\lim_{t \to \infty} d \phi(t) / d t = 0$,  we obtain from 
Eq.~\equ{eq:generalphi} the following self-consistent equation 
for $\Fi$
\begin{equation}
    \phi_\infty  = 1  -  \frac{1}{\avk} \sum_k k P(k) e^{ -\lambda k \phi_\infty}.
\end{equation}
The value $ \phi_\infty = 0$ is always a solution. In order to have a non-zero
solution, the condition 
\begin{equation}
    \frac{d}{d \phi_\infty } \left. \left( 1  -  \frac{1}{\avk} \sum_k k P(k) e^{
          -\lambda  k \phi_\infty} \right) \right|_{\phi_\infty=0} > 1
\end{equation}
must be fulfilled. This relation implies
\begin{equation}
 \frac{1}{\avk} \sum_k k P(k) (\lambda k) = \lambda \frac{\fluck}{\avk} > 1.
\end{equation}
This condition defines the epidemic threshold
\begin{equation}
  \lambda_c =  \frac{\avk}{\fluck}
  \label{eq:threshold}
\end{equation}
below which the epidemic incidence is null, and above which it 
attains a finite value. That is, the threshold 
is inversely proportional to the connectivity
fluctuations $\fluck$. For regular networks, in which $\fluck<\infty $, the
threshold has a finite value and we are in the presence of a standard
phase transition. 
On the other hand, networks with  strongly
fluctuating connectivity distribution, show 
a {\em vanishing} epidemic threshold for increasing network sizes; i.e.
$\fluck\to\infty $ for $N\to \infty$.
The absence of any intrinsic epidemic threshold in this network can 
be understood by noticing that in
heterogeneous systems the basic reproductive number ${\cal R}_0$
contains a correction term linearly dependent on the fluctuations
(standard deviation) of the nodes' connectivity distribution
\cite{anderson92,lloyd01}. In SF networks the divergence of the
connectivity fluctuations leads to an ${\cal R}_0$ that always exceeds
unity whatever the spreading rate $\lambda$ is. This ensures that epidemics
always have a finite probability to survive indefinitely.  It is worth
remarking that real networks have always a finite size $N$ and thus an
effective threshold, depending on the magnitude of $\avk$ and
$\fluck$, that can be easily calculated as a function of $N$. This
apparent threshold, however is not an intrinsic quantity and it is
extremely small for systems with large enough $N$.

\section{Exponentially distributed networks: The Watts-Strogatz model}
\label{sec:exponential}

The class of exponential networks refers to random graph models which
produce a connectivity distribution $P(k)$ peaked at an average value
$\left<k\right>$ and decaying exponentially fast for $k \gg
\left<k\right>$ and $k\ll\left<k\right>$.  Typical examples of such a
network are the random graph model \cite{erdos60} and the small-world
model of Watts and Strogatz (WS) \cite{watts98}. The latter has
recently been the object of several studies as a good candidate for
the modeling of many realistic situations in the context of social and
natural networks. In particular, the WS model shows the
``small-world'' property common in random graphs \cite{watts99}; i.e.,
the diameter of the graph---the shortest chain of links connecting
any two vertices---increases very slowly, in general logarithmically
with the network size \cite{alain}.  On the other hand, the WS model
has also a local structure (clustering property) that is not found in 
random graphs with finite connectivity \cite{watts98,alain}.  The WS
graph is defined as follows \cite{watts98,alain}: The starting point
is a ring with $N$ nodes, in which each node is symmetrically
connected with its $2m$ nearest neighbors.  Then, for every node each
link connected to a clockwise neighbor is rewired to a randomly chosen
node with probability $p$, and preserved with probability $1-p$. This
procedure generates a random graph with a connectivity distributed
exponentially for large $k$ \cite{watts98,alain}, and an average
connectivity $\left<k \right> = 2 m$. The graph has small-world
properties and a non-trivial ``clustering coefficient''; i.e.,
neighboring nodes share many common neighbors \cite{watts98,alain}.
The richness of this model has stimulated an intense activity aimed at
understanding the network's properties upon changing $p$ and the
network size $N$~\cite{watts99,nw99,watts98,alain,amaral2,bra99}. At
the same time, the behavior of physical models on WS graphs has been
investigated, including epidemiological percolation models
\cite{newman00,nw99,moore00} and models with epidemic cycles
\cite{abramson01}.

In the following we focus on the WS model with $p=1$; it is 
worth noticing that even in this extreme case the network 
retains some memory of the generating procedure. 
The network, in fact, is not locally equivalent
to a random graph in that each node has at least $K$ neighbors. 
In the limit $p\to1$, the connectivity distribution of the WS network,
as defined previously, takes the form \cite{barrat00}
\begin{equation}
  \label{eq:wspk}
  P(k) = \frac{m^{k-m}}{(k-m)!} e^{-m} \qquad {\rm for }\;\; k \geq m.
\end{equation}
This is a Poisson distribution, with finite moments. Defining the
{\em factorial moments} \cite{gardiner}
\begin{equation}
  \left< X^r \right>_f \equiv  \left< X (X-1)(X-2) \cdots (X-r+1) \right>,
\end{equation}
we have for the distribution \equ{eq:wspk}
\begin{equation}
  \left< (k-m)^r \right>_f = m^r.
  \label{eq:momenta}
\end{equation}
In particular, from Eq.~(\ref{eq:momenta}), the first  moments of the
connectivity distribution are given by
\begin{eqnarray}
  \left< k\right> & = & 2m,\\
   \left< k^2 \right> & = & m(1+4m),\\
   \left< k^3 \right> & = & m(1+6m+8m^2),
\end{eqnarray}
and, in general, 
\begin{equation}
  \left< k^n \right> \sim (2 m)^n,
\end{equation}
for large $m$.

For general regular networks, for which $\left< k^n \right> < \infty$ for
all values of $n$, Eqs.~\equ{eq:generalphi} and \equ{eq:rigeneral} can
be approximately solved in the limit $ \phi(t) \to 0$, by expanding the
exponentials under the summation signs. Thus, for the case of the
total epidemic incidence $\Ri$ in Eq.~\equ{eq:rigeneral},
\begin{equation}
  \Ri \simeq  \sum_k P(k) \lambda k \phi_\infty = \avk \lambda  \phi_\infty.
  \label{eq:Rigennet}
\end{equation}
That is, $\Ri$  is linearly proportional to 
$\phi_\infty$. 

On the other hand, by expanding the
exponential in Eq.~\equ{eq:generalphi} and keeping the most relevant 
terms, we yield :
\begin{equation}
  \frac{ d \phi}{d t}
  \simeq  1 - \phi -  \frac{ \sum_k k P(k) (1 -
    \lambda k \phi + \lambda^2 k^2 \phi^2 /2)}{\avk} 
  = \phi \left(-1 + \lambda \frac{\fluck}{\avk} -  \lambda^2 \phi \frac{\left< k^3
      \right>}{2 \avk} \right).
\end{equation}
The resulting previous equation can be exactly solved, to yield
\begin{equation}
  \phi(t) = \frac{2(\lambda-\lambda_c)}{\fluck} \frac{1}{\left< k^3 \right> \lambda^2+ A
    e^{-(\lambda-\lambda_c) t / \lambda_c}},
\end{equation}
where $\lambda_c$ is defined as in Eq.~\equ{eq:threshold}.
That is, for $\lambda < \lambda_c$, $\phi_\infty \to 0$, while from $\lambda > \lambda_c$, we recover
the well-known mean-field 
behavior $\phi_\infty \sim (\lambda-\lambda_c)$, which translated
to the total epidemic incidence $\Ri$ yields
\begin{equation}
  \Ri \sim (\lambda-\lambda_c)
  \label{eq:riexpo}
\end{equation}

\begin{figure}[t]
  \centerline{\epsfig{file=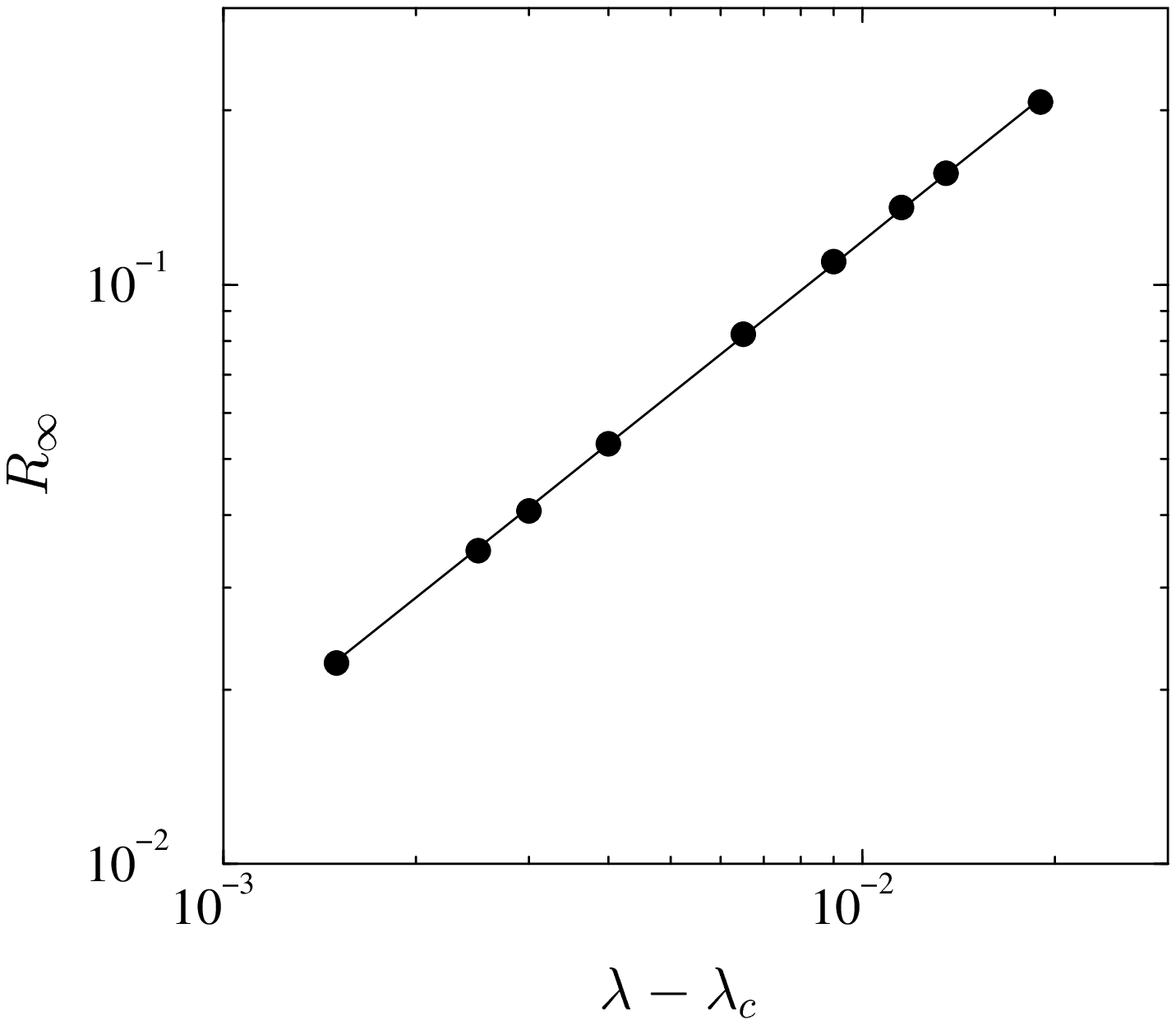, width=9cm}} 
  \vspace*{0.5cm}
  \caption{Total density of infected individuals $\Ri$ as a function of
    $\lambda-\lambda_c$ for the SIR model in WS networks of size $N=10^6$ . 
    The value of $\lambda_c=0.184(5)$ is in good agreement with the analytical
        prediction. The full line is a fit to the form 
     $R_\infty\sim (\lambda-\lambda_c)^\beta$ with $\beta=0.9(1)$.}
\label{fig1}

\end{figure}

\begin{figure}[t]
  \centerline{\epsfig{file=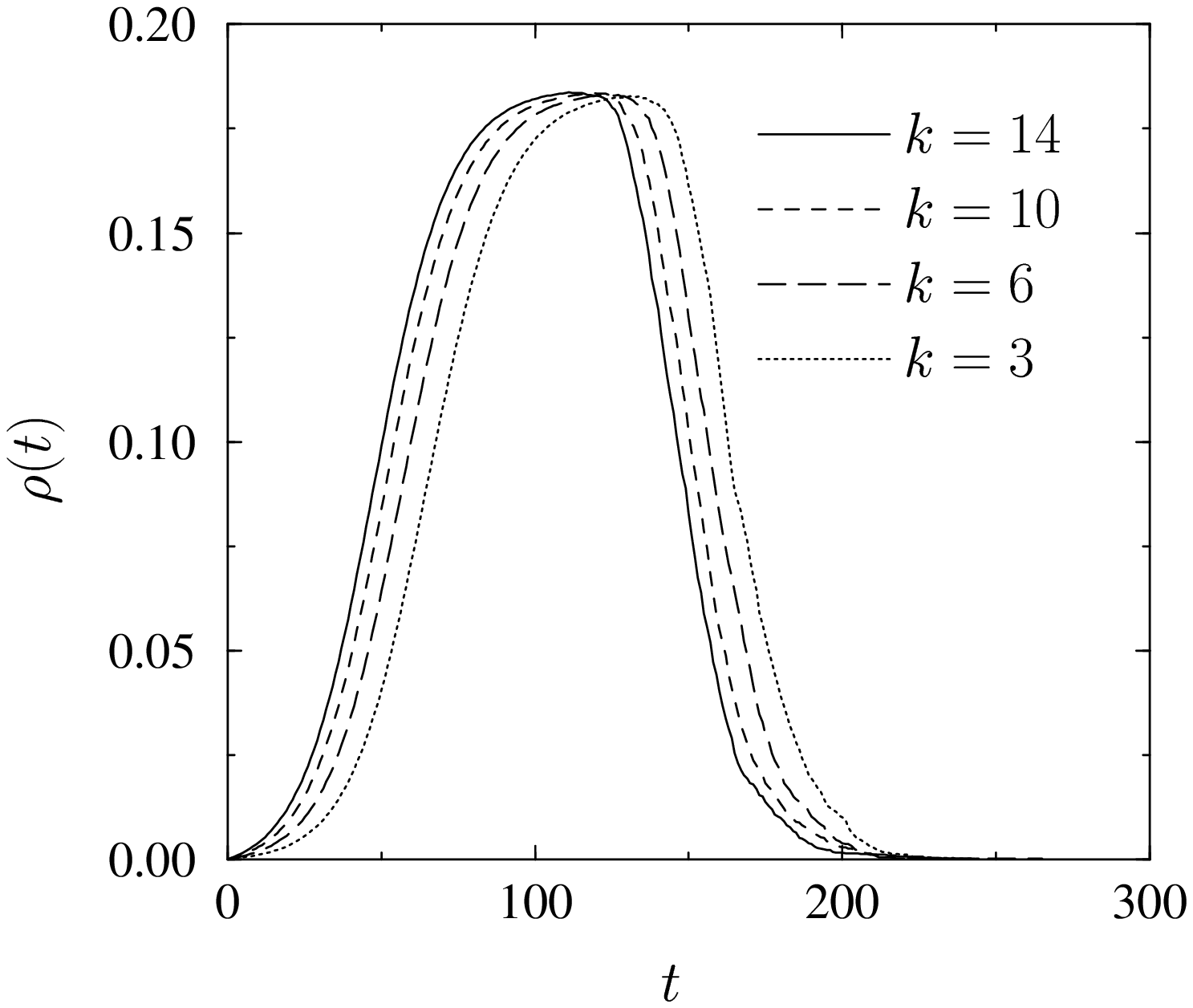, width=9cm}} 
  \vspace*{0.5cm}
  \caption{Total density of infected nodes as a function
    of time for the SIR model in  WS networks, starting from initial
    conditions peaked in nodes of connectivity $k$ (initial
    infected individuals randomly distributed only among the nodes of
    connectivity $k$). The spreading rate is fixed to $\lambda=0.20$. The
    network size is $N=10^6$.}
\label{fig2}

\end{figure}

In the particular case of the WS networks, we expect to observe the
behavior dictated by Eq.~\equ{eq:riexpo}, with an epidemic threshold given by
\begin{equation}
  \lambda_c =  \frac{\avk}{\fluck}=\frac{2}{1+4m}.
\end{equation}
In order to compare with the analytical predictions we have carried
out large scale simulations of the SIR model in the WS network with
$p=1$. In our simulations we consider the WS network with parameter
$m=3$, which corresponds to an average connectivity $\left< k \right>
= 6$. Simulations were implemented on graphs with number of nodes
ranging from $N=10^3$ to $N=3 \times 10^6$, averaging over at least $10^4$
different epidemic outbreaks, performed on at least $10$ different
realization of the random network. In Fig.~\ref{fig1}, we show the
total density of removed nodes at the end of the epidemic outbreak as
a function of the parameter $\lambda$. The graph exhibits an epidemic
threshold at $\lambda_c=0.184(5)$ that is approached with a 
roughly linear behavior by
$R_\infty$. A linear fit to the form 
$R_\infty \sim (\lambda-\lambda_c)^\beta$ provides an exponent
$\beta=0.9(1)$, in reasonable agreement with the analytical finding.  This
confirms that the SIR model in exponentially bounded complex networks
has a behavior similar to that obtained with the homogeneous mixing
hypothesis. Actually, since the connectivity fluctuations are very
small in the WS graph ($\left< k^2 \right> \sim\left<k\right>$), as a
first approximation we can consider the WS model as a homogeneous
one in which each node has the same number of links, $k\simeq \left< k
\right>$.  In order to provide further evidence to this effective
homogeneity, we show in Fig.~\ref{fig2} the time evolution of the
density of infected nodes for epidemic outbreaks starting only on
nodes with a given connectivity $k$. The total epidemic 
incidence is almost constant for all connectivity $k$, 
with a slight shift of the peak time of the outbreak.
The figure clearly shows that the
system reacts almost identically to this heterogeneous initial
conditions, confirming that the homogeneity assumption is correctly
depicting the system's behavior.  We shall see in the next section
that this is not the case for SF networks.

\section{Power-law distributed networks: The Barab\'{a}si-Albert model}
\label{seq:BA}
  
  The Barab\'{a}si-Albert (BA) graph was introduced as a model of
  growing network (such as the Internet or the world-wide-web) in
  which the successively added nodes establish links with higher
  probability pointing to already highly connected nodes
  \cite{barab99}.  This is a rather intuitive phenomenon on the
  Internet and other social networks, in which new individuals tend to
  develop more easily connections with individuals which are already
  well-known and widely connected.  The BA graph is constructed using
  the following algorithm \cite{barab99}: We start from a small number
  $m_0$ of disconnected nodes; every time step a new vertex is added,
  with $m$ links that are connected to an old node $i$ with
  probability
\begin{equation}
  \Pi(k_i) = \frac{k_i}{\sum_j k_j},
\end{equation}
where $k_i$ is the connectivity of the $i$-th node.  After iterating
this scheme a sufficient number of times, we obtain a network composed
by $N$ nodes with connectivity distribution $P(k) \sim k^{-3}$ and
average connectivity $\left<k \right> = 2 m$ (in this work we will
consider the parameters $m_0=5$ and $m=3$). Despite the well-defined
average connectivity, the scale invariant properties 
turn out to play a major role on the physical properties of 
these networks (for instance, the resilience 
to attack \cite{newman00,havlin01}).

In the continuous $k$ approximation, that substitutes the discrete
variable $k$ for a continuous variable in the range $[m, \infty[$, the
connectivity distribution of the BA model takes the form
\begin{equation}
  P(k) = \frac{2 m^2}{k^3} \qquad {\rm for }\;\; k \geq m.
  \label{eq:condisBA}
\end{equation}
For this distribution, the first moment is finite, $\left<k \right> =
2 m$, but the second moment diverges with the network size, $\fluck \sim
\log N$. In view of the general result, Eq.~(\ref{eq:threshold}), we
observe that the epidemic threshold in this particular network tends
to zero for large $N$. Also, the general solutions obtained in
Sec.~4 cannot be applied, and we must work out
the particular solutions of Eqs.~(\ref{eq:rigeneral})
and~(\ref{eq:generalphi}). 

The equation for $\Ri$, with the connectivity distribution
(\ref{eq:condisBA}) is
\begin{equation}
  \Ri = 1 - 2 m^2 \int_m^\infty k^{-3} e^{-\lambda k \Fi} dx
  = 1 - 2 z^2 \int_z^\infty x^{-3} e^{-x} dx,
\end{equation}
where we have defined the new variable $z=\lambda m \phi$. Performing the
integral, we obtain
\begin{equation}
  \Ri = 1 - e^{-z}(1-z) - z^2 \Gamma(0,z),
\end{equation}
where $\Gamma(a,z)$ is the incomplete Gamma function \cite{abramovitz}. 
For small
values of $z$, the function $\Gamma(0,z)$ can be expanded in the expression
\cite{abramovitz}
\begin{equation}
   \Gamma(0,z) \simeq -(\gamma_E + \ln(z)) + z + {\cal O}(z^2),
\end{equation}
where $\gamma_E$ is the Euler's constant. By inserting this expansion into the
expression for $\Ri$, we obtain for small
values of $\Fi$
\begin{equation}
  \Ri \simeq 2 z  \equiv \lambda \avk \Fi.
\end{equation}

On its turn, the equation for $\phi(t)$, with the connectivity
distribution (\ref{eq:condisBA}) is
\begin{equation}
  \frac{ d \phi(t)}{d t} = 1 - \phi(t) -  m  \int_m^\infty k^{-2} e^{-\lambda k \phi} dk.
\end{equation}
Defining the new variable $z=\lambda m \phi$, we can rewrite the previous
equation as 
\begin{equation}
  \frac{1}{\lambda m}\frac{d z}{d t} = 1 - \frac{z}{\lambda m} -  z  \int_z^\infty x^{-2}
  e^{-x} dx. 
  \label{eq:BAphi}
\end{equation}
In order to study the limit $z (\phi) \to 0$ we must first integrate by
parts the integral in Eq.~(\ref{eq:BAphi}), to get
\begin{equation}
  \frac{1}{\lambda m}\frac{d z}{d t} =  1 - \frac{z}{\lambda m} -  e^{-z} - z
  \ln(z) e^{-z} - z  \int_z^\infty \ln(x) e^{-x} dx.
  \label{eq:BAphi2}
\end{equation}
We can now take the limit $z\to0$ in the last integral, obtaining

\begin{equation}
  \int_0^\infty \ln(x) e^{-x} dx = - \gamma_E,
\end{equation}
where again $\gamma_E$ is Euler's constant. Introducing this approximation,
and Taylor expanding the expression (\ref{eq:BAphi2}), we obtain 
\begin{equation}
  \frac{1}{\lambda m}\frac{d z}{d t} \simeq  z\left[1 -\gamma_E - \frac{1}{\lambda m} -
    \ln(z) \right].
\end{equation}
This equation can be integrated, to yield
\begin{equation}
  \phi\simeq  \frac{1}{\lambda m}  \exp \left( 1 -\gamma_E - \frac{1}{\lambda m} - A e^{-\lambda m t}
  \right),
\end{equation}
where $A$ is an integration constant. The stationary regime for long
times is 
\begin{equation}
  \phi_\infty \simeq  \frac{e^{1-\gamma_E}}{\lambda m}  e^{-1/\lambda m},
\end{equation}
and by inserting this result into the expression 
for the total epidemic incidence we find 
\begin{equation}
  \Ri \sim   e^{-1/\lambda m}.
  \label{eq:stretchBA}
\end{equation}
That is, the function $\Ri$ is non-zero for any 
non-zero value of $\lambda$,
which is in agreement with the predicted 
threshold $\lambda_c=0$. This result also  recovers the same behavior 
obtained by considering a diverging connectivity variance 
in the results reported by May and Anderson for HIV spreading
in heterogeneous population\cite{may88,lloyd01}.

\begin{figure}[t]
  \centerline{\epsfig{file=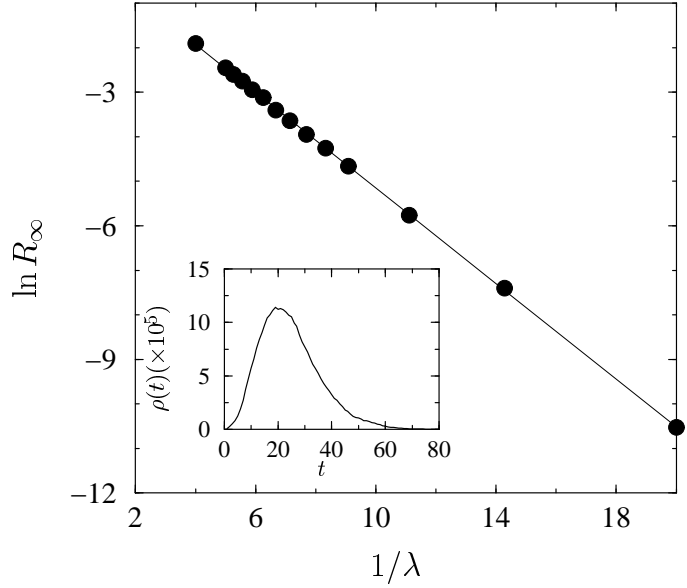, width=9cm}} 
  \vspace*{0.5cm}
  \caption{Total density  of infected individuals $R_\infty$ as a function of
    $1/ \lambda$ for the SIR model in BA networks of size $N=10^6$ .
    The linear behavior on the semi-logarithmic scale proves the
    stretched exponential behavior predicted by
    Eq.~\equ{eq:stretchBA}. The inset show the time profile of 
    the average density of infected individuals at the 
    spreading rate $\lambda=0.9$.}
\label{fig3}

\end{figure}

\begin{figure}[t]
  \centerline{\epsfig{file=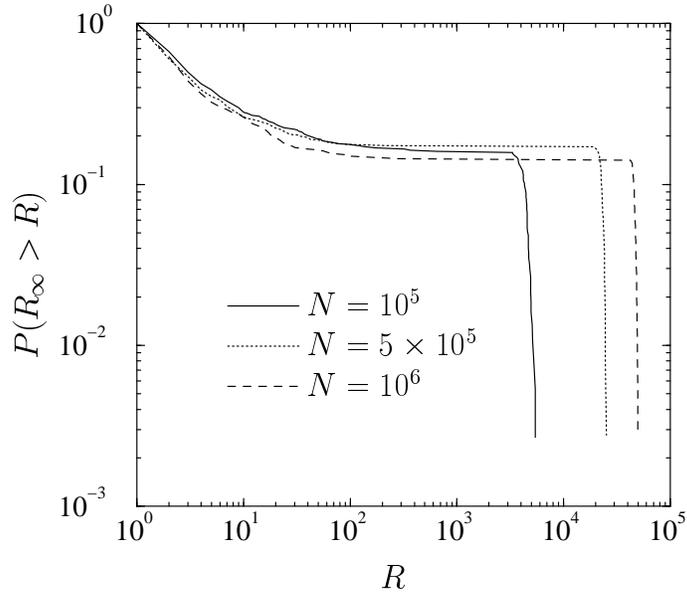, width=9cm}} 
  \vspace*{0.5cm}
  \caption{Cumulated outbreak epidemic distribution for the SIR model
    in BA networks. The spreading rate is fixed to $\lambda=0.09$.} 
\label{fig4}

\end{figure}

The numerical simulations performed on the BA network confirm the
picture extracted from the analytic treatment. We consider 
the SIR model on BA networks of
size ranging from $N=10^3$ to $N=10^6$, with $m=3$ and thus $\avk=6$.
As predicted by the analytic calculations,  
Fig.~\ref{fig3} shows that $R_\infty$ 
decays with  $\lambda$ as $R_\infty\sim \exp(-C/\lambda)$, 
where $C$ is a constant. In order to rule out
the presence of finite size effects hiding an abrupt transition (the
so-called smoothing out of critical points \cite{marro99}), we have
inspected the behavior of the stationary persistence for network sizes
varying over three orders of magnitude.
The total absence of scaling
of $R_\infty$ and the perfect agreement for any size with the analytically
predicted exponential behavior allows us to definitely confirm the
absence of any finite epidemic threshold.
A closer look at $R_\infty$ is given in Fig.~\ref{fig4}.
While Fig.~\ref{fig3} reports the average over $10^4-10^5$ epidemic 
outbreaks, Fig.~\ref{fig4} reports an illustration of the behavior 
of the cumulative probability $P(R_\infty>R)$ of having an outbreak 
which affects $R$ individuals in a single realization 
at $\lambda=0.09$. The figure shows a finite probability of 
having outbreaks involving a number of individuals of the order 
of the network size $N$. The large plateau corresponds 
to a gap between large events and small outbreaks that 
give rise to a zero density of infected individuals in the 
$N\to \infty$ limit. Accordingly to the predictions, the 
plateau extends proportionally to $N$ for increasing network sizes. 

\begin{figure}[t]
  \centerline{\epsfig{file=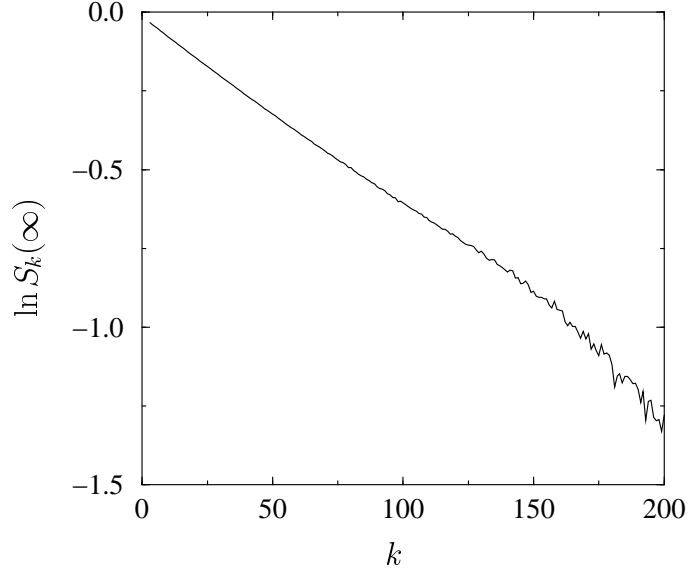, width=9cm}} 
  \vspace*{0.5cm}
  \caption{Density $S_k(\infty)$ of susceptible nodes as a function
    of $k$  for the SIR model in BA networks. Epidemics start 
    from random initial conditions (initial infected individuals randomly
    distributed among all nodes). The spreading rate is fixed to
    $\lambda=0.09$. The network size is $N=10^6$. The linear-log plot
    recovers the exponential form predicted in Eq.~\equ{eq:4} }
\label{fig5}
\end{figure}

\begin{figure}[t]
  \centerline{\epsfig{file=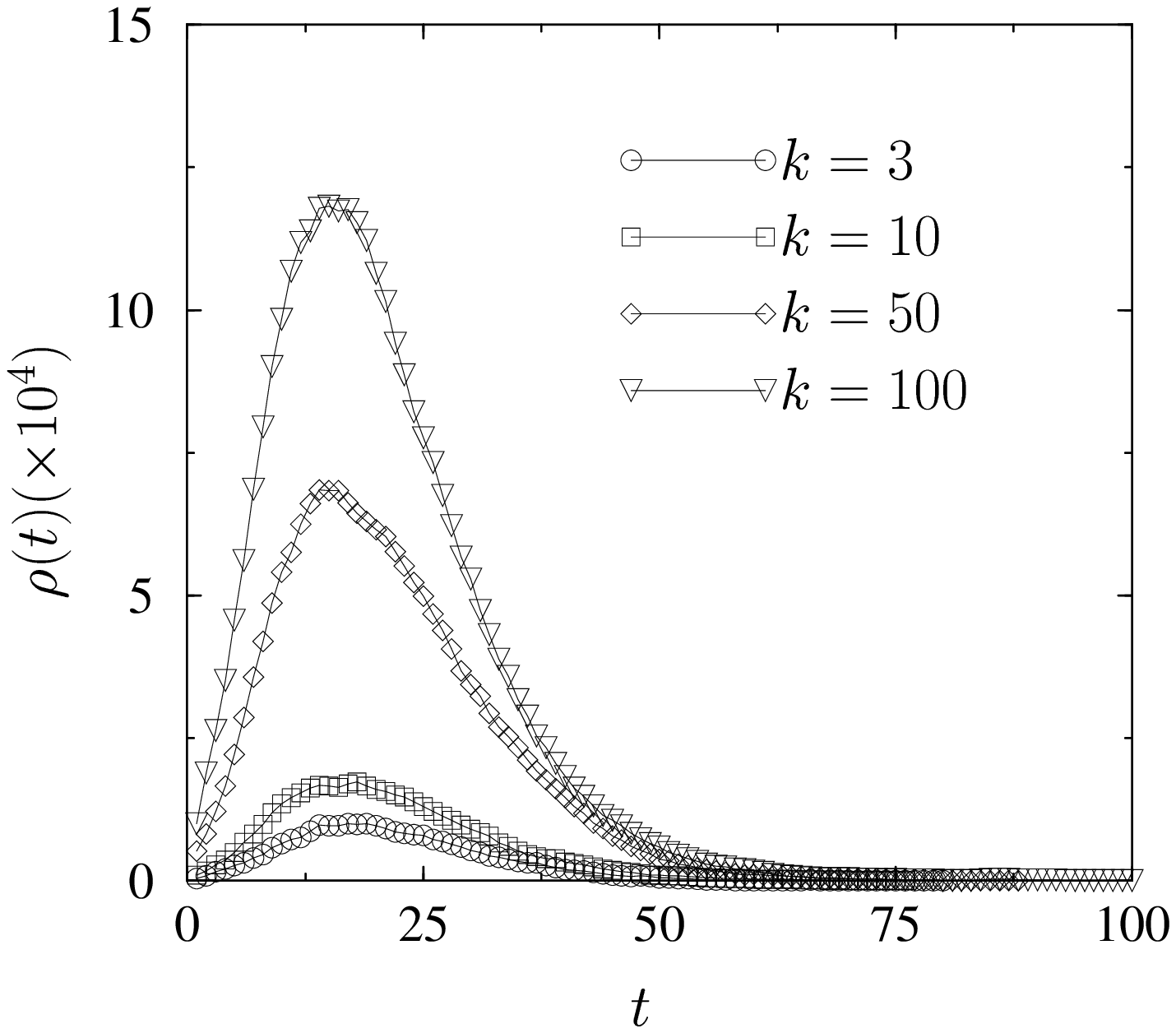, width=9cm}} 
  \vspace*{0.5cm}
  \caption{Total density of infected nodes as a function
    of time for the SIR model in BA networks, starting from initial
    conditions concentrated in nodes of connectivity $k$ (initial
    infected individuals randomly distributed among the nodes of
    connectivity $k$). The spreading rate is fixed to $\lambda=0.09$. The
    network size is $N=10^6$.}
\label{fig6}

\end{figure}

The large heterogeneity of these networks can be pictorially
characterized by inspecting the epidemic evolution in each class of
connectivity $k$. We know from Eq.~(\ref{eq:4}), that the susceptibles
densities $S_k(t)$ decay much faster in the highly connected classes.
In particular, we have that $S_k(\infty)\sim \exp(-\lambda k \phi_\infty)$. In
Fig.~\ref{fig5}, we report $S_k(\infty)$ as a function of $k$ on a
semi-logarithmic scale.  The plot shows the expected linear relation
in $k$. Since $R_k(\infty)=1-S_k(\infty)$, the curves clearly show that the
higher is the nodes' connectivity, the higher is the relative
incidence of the epidemic outbreak. Classes of nodes with few
connections have a small density of removed individuals (total number
of infected individuals), while highly connected classes ($k\gg 100$)
are almost totally affected by the infection.  A further striking
evidence of the peculiar behavior of the SF networks is obtained by
inspecting epidemic outbreaks starting on nodes with different
connectivity $k$. While an analytical solution for this case is very
troublesome, the numerical investigation presents clear-cut results.
In Fig.~\ref{fig6} we present the infection incidence profile for
epidemic outbreaks started on sites with different connectivities $k$.
The population results much weaker (higher number of infected
individuals) to epidemics starting on highly connected individuals.
This weakness points out that the best protection of these networks
can be achieved by targeted immunization programs
\cite{anderson92,psvpro}.

\section{Generalized scale-free networks}

Recently there has been a burst of activity in the modeling of SF
complex networks. The recipe of Barab\'{a}si and Albert \cite{barab99} has
been followed by several variations and generalizations
\cite{albert00,mendes99,krap00,bosa00} and the revamping of previous
mathematical works \cite{simon55}.  All these studies propose methods
to generate SF networks with variable exponent $\gamma$.  The
analytical treatment presented in the previous section for the SIR
model can be easily generalized to SF networks with connectivity
distribution with $\gamma>0$. Let us consider a generalized SF network with a
normalized connectivity distribution given by
\begin{equation}
  P(k) = (1+\gamma) m^{1+\gamma}  k^{-2-\gamma},
  \label{eq:connectgen}
\end{equation}
where we are approximating the connectivity $k$ as a continuous
variable and assuming $m$ the minimum connectivity of any node.
The average connectivity is thus
\begin{equation}
  \left< k \right> = \int_m^\infty k P(k) dk = 
  \frac{1+\gamma}{\gamma}  m,    
\end{equation}
while the connectivity fluctuations are given by
\begin{eqnarray}
  \fluck &=& \frac{\gamma+1}{\gamma-1} m^2 \qquad  {\rm if} \;\; \gamma>1,\\
  \fluck &=&\infty \qquad \qquad \;\; \; {\rm if} \;\; \gamma\leq 1.
\end{eqnarray}
Thus, according to the general result Eq.~(\ref{eq:threshold}), the
epidemic threshold, as a function of $\gamma$ is 
\begin{eqnarray}
  \lambda_c &=& \frac{\gamma-1}{\gamma m}  \qquad  {\rm if} \;\; \gamma>1,
  \label{eq:thresholdgamma}\\ 
  \lambda_c &=&0 \qquad \qquad  {\rm if} \;\; \gamma\leq 1.
\end{eqnarray}

To obtain the explicit expression for $\Fi$ and $\Ri$ we must solve the
Eqs.~(\ref{eq:generalphi}) and~(\ref{eq:rigeneral}) for the general
connectivity distribution (\ref{eq:connectgen}). While the
differential equation~(\ref{eq:generalphi}) cannot be solved in a
closed form for general $\gamma$, we can obtain approximation to the steady
state value at long times, $\Fi$, solving the algebraic equation 
\begin{equation}
  \Fi = 1 -   \frac{1}{\avk} \sum_k k P(k) e^{ -\lambda k  \Fi }. 
  \label{eq:figennet}
\end{equation}
As a function of $\Fi$, from Eq.~(\ref{eq:rigeneral}), 
the total epidemic incidence $\Ri$ takes the form, 

\begin{eqnarray}
1-\Ri &=& (1+\gamma)m^{1+\gamma} \int_m^\infty dk k^{-2-\gamma} e^{-\lambda k \Fi}
  = (1+\gamma) z^{1+\gamma} \int_z^\infty x^{-2-\gamma} e^{-x} dx \\
  &=& (1+\gamma) z^{1+\gamma} \Gamma(-1-\gamma, z),
\end{eqnarray}
where we have defined $z=\lambda k \Fi$ and $\Gamma(a,z)$ is the incomplete Gamma
function \cite{abramovitz}.  In the limit $z\to0$, we can perform a
Taylor expansion of the incomplete Gamma function, with the form
\cite{abramovitz} 
\begin{equation}
  \Gamma(a,z) = \Gamma(a) - \frac{z^a}{a} + z^a \sum_{n=1}^\infty \frac{(-z)^n}{(a+n) n!},
\end{equation}
where $\Gamma(a)$ is the standard Gamma function. This expansion has
obviously meaning only for $a\neq -1, -2, -3, \ldots$ Thus, integer values of
$\gamma$ must be analyzed in a case by case basis.  Substituting the last
formula into the expression for $\Ri$, we are led to
\begin{equation}
  \Ri = \Gamma(-\gamma) z^{1+\gamma} - (1+\gamma) \sum_{n=1}^\infty \frac{(-z)^n}{(n-\gamma-1) n!} 
  \simeq  \frac{\gamma+1}{\gamma} z + {\cal O} (z^{1+\gamma}).
\end{equation}
That is, for any value of $\gamma>0$, the leading behavior of $\Ri$ is
\begin{equation}
  \Ri \sim \frac{\gamma+1}{\gamma} z = \lambda \avk \Fi,
  \label{eq:Ridef}
\end{equation}
which is equivalent to the expression found for regular networks in
Eq.~(\ref{eq:Rigennet}). 

In order to find the infinite time limit  value $\Fi$, we must solve the
Eq.~(\ref{eq:figennet}). Substituting the form of the generalized
connectivity distribution~(\ref{eq:connectgen}), we have
\begin{equation}
  \Fi = 1 - \gamma m^\gamma \int_m^\infty k^{-1-\gamma} e^{-\lambda \Fi k} dk
  = 1 - \gamma z^\gamma \int_z^\infty x^{-1-\gamma} e^{-x} dx 
  = 1 - \gamma z^\gamma \Gamma(-\gamma, z)
\end{equation}
where again $z=\lambda k \Fi$. Inserting in this last expression the Taylor
expansion for the incomplete Gamma function, we obtain
\begin{equation}
  \Fi = z^\gamma \Gamma(1-\gamma) + \gamma  \sum_{n=1}^\infty \frac{(-z)^n}{(n-\gamma) n!}
  \label{eq:TaylorFi}
\end{equation}
The leading behavior of the r.h.s. of this equation depends on the
particular value of $\gamma$ considered.

(a) $0<\gamma<1$: In this case, we have
\begin{equation}
  \Fi \simeq  (\lambda m \Fi)^\gamma \Gamma(1-\gamma),
\end{equation}
from which we obtain 
\begin{equation}
  \Fi \simeq  (\lambda m)^{\gamma/(1-\gamma)} \Gamma(1-\gamma)^{1/(1-\gamma)}. 
\end{equation}
Combining this result with Eq.~(\ref{eq:Ridef}), we obtain
\begin{equation}
  \Ri \sim   \lambda^{1/(1-\gamma)}.
\end{equation}
In this range of values of $\gamma$ we recover the absence of the epidemic
threshold, and the associated critical behavior, as we have already
shown in Sec.~\ref{seq:BA}.

(b) $1<\gamma<2$: To obtain a nontrivial solution for $\Fi$, we must keep
the first two terms in the Taylor expansion~(\ref{eq:TaylorFi}),
namely:
\begin{equation}
  \Fi \simeq  (\lambda m \Fi)^\gamma \Gamma(1-\gamma) + \frac{\gamma}{\gamma-1} \lambda m \Fi.
\end{equation}
From this equation we find
\begin{equation}
  \Fi \simeq  \left[ \frac{\gamma}{\Gamma(2-\gamma)} \frac{m}{(\lambda m)^\gamma} \left( \lambda
      -\frac{\gamma-1}{\gamma m}\right)  \right]^{1/(\gamma-1)},
\end{equation}
which yields 
\begin{equation}
  \Ri \simeq  (\lambda -\lambda_c)^{1/(\gamma-1)},
\end{equation}
with an epidemic threshold $\lambda_c$ given by Eq.~(\ref{eq:thresholdgamma}). 

(c) $\gamma>2$: The most relevant terms in the expansion of $\Fi$ are now 
\begin{equation}
  \Fi \simeq  \gamma \frac{\lambda m \Fi}{\gamma-1} - \gamma \frac{(\lambda m \Fi)^2}{\gamma-2}.
\end{equation}
The relevant expression for $\Fi$ is
\begin{equation}
  \Fi \simeq  \frac{\gamma-2}{\gamma-1} \frac{1}{\lambda^2 m} \left( \lambda
      -\frac{\gamma-1}{\gamma m}\right),
\end{equation}
that for the epidemic incidence yields the behavior 
\begin{equation}
  \Ri \simeq  (\lambda -\lambda_c).
\end{equation}
The threshold $\lambda_c$ is again given by the general
expression~(\ref{eq:thresholdgamma}).
In other words, we recover the usual epidemic
framework in networks with connectivity distribution that decays
faster than $k$ to the fourth power. Obviously, an exponentially
bounded network is  included in this last case.

\section{Conclusions}

The presented results for the SIR model in complex networks
confirm the epidemiological picture proposed in previous works.  
The topology of the network has a great influence in the overall 
behavior of epidemic spreading. The connectivity fluctuations
of the network play a major role by strongly enhancing 
the infection's incidence. 
This issue assumes a particular relevance in the case of 
SF networks that exhibit connectivity fluctuations diverging 
with the increasing size $N$ of the web.
SF networks are therefore  very weak in face of infections 
presenting an effective epidemic threshold that is vanishing
in the limit $N\to \infty$.
In the case of the SIR model in an infinite population 
this corresponds to the absence 
of any epidemic threshold below which major epidemic outbreaks are 
impossible. These results strengthen the epidemiological framework 
for complex networks reported for the SIS model \cite{pv01a,pv01b}
and proposed as well for the SIR model \cite{lloyd01}. 
The emerging picture is likely going to stimulate the re-analysis
of several concepts of standard epidemiology such as the 
``core group'' or the characteristic number of 
contacts that appears to be ill-defined in SF networks. 

The high heterogeneity of SF networks finds signatures
also in the peculiar susceptibility  
to infections starting on the most connected individuals 
and the different relative incidence within populations of
varying connectivity $k$. It is reasonable to expect that 
these features can point at better protection methods for 
these networks which appear to have practical realization 
in many technological and biological systems. 
In this perspective, the introductions 
of many elements of realism and a better knowledge of 
the networks temporal pattern are  fundamental ingredients
towards a better understanding of the spreading of information 
and epidemics in a wide range of complex interacting systems.\\

{\small This work has been partially supported by the European Network
  Contract No. ERBFMRXCT980183. RP-S also acknowledges 
  support from the grant CICYT PB97-0693.
  We thank M.-C.  Miguel, R. V.  Sol{\'e},
  S. Visintin, and R.  Zecchina for helpful
  comments and discussions. We are grateful to A.~L. Lloyd 
  and R.~M.~May for 
  enlightening suggestions and for pointing out to us some 
  fundamental references.}


\newpage

\end{document}